\documentclass{optica-article}
\journal{oe}
\articletype{Research Article}

\usepackage{siunitx}
\usepackage{lineno}
\DeclareSIUnit{\rad}{rad}
\newcommand\sminus{\scalebox{1}[0.99]{$-$}}

\begin{document}

\title{Tilting refractive x-ray lenses for fine-tuning of their focal length}

\author{Rafael Celestre\authormark{1,*}, Thomas Roth\authormark{1}, Carsten Detlefs\authormark{1},\\ Peng Qi\authormark{2}, Marco Cammarata\authormark{1}, Manuel Sanchez del Rio\authormark{1}, Raymond Barrett\authormark{1}}
\address{\authormark{1}The European Synchrotron, 71 Avenue des Martyrs, 38000 Grenoble, France\\
\authormark{2}Paul Scherrer Institut, Forschungsstrasse 111, 5232 Villigen, Switzerland}
\email{\authormark{*}rafael[dot]celestre[at]]esrf[dot]eu}

\begin{abstract}
In this work, we measure and model tilted x-ray refractive lenses to investigate their effects on an x-ray beam. The modelling is benchmarked against at-wavelength metrology obtained with x-ray speckle vector tracking experiments (XSVT) at the BM05 beamline at the ESRF-EBS light source, showing very good agreement. This validation permits us to explore possible applications of tilted x-ray lenses in optical design: we demonstrate that tilting 1D lenses around their focusing direction can be used for fine-tuning their focal length with possible applications in beamline optical design.
\end{abstract}

\section{Introduction}\label{sec:intro}

After some initial controversy \cite{Suehiro1991, Michette1991}, x-ray lenses were established as a viable optical element in the mid-1990s \cite{Tomie1994, Snigirev1996}, becoming common in x-ray beamlines within a few years of their first experimental demonstration~\cite{Lengeler2001}. In this work, we concentrate on the yet-to-be-explored field of tilted refractive x-ray lenses.

Regarding applications of tilted optical elements, curved x-ray mirrors have a focal length that is a function of incident angle and their radius of curvature - see Coddington's equations for x-ray mirrors in \cite{Howells1993}. Rotating diffractive x-ray lenses was investigated by David \textit{et al.}~\cite{David2001}, where they tilted 1D zone plates to make use of the increased projected thickness along the optical axis to improve their diffraction efficiency. Tilting of multi-layer Laue lenses (MLL) has also been reported as a way of increasing their diffraction efficiency \cite{Yan2010}. Lastly, x-ray prism arrays arranged to form a focusing device (i.e.~``alligator lens'') \cite{Cederstrom2000} can have their focal length varied by rotation of the individual prism arrays relative to each other as demonstrated in \cite{Jark2019}. Yet, applications of tilted parabolic refractive x-ray lenses are not found in literature. 

Despite the effort in simulating ideal and aberrated lenses \cite{Alianelli2007, Baltser2011, SanchezdelRio2012, Osterhoff2017, celestre_modelling_2020}, little to no attention has been given to the modelling of tilted refractive x-ray lenses. One of the scarce works on this topic was presented by Andrejczuk \textit{et al}.~\cite{Andrejczuk2010}. Their work discusses the effects of tilting as observed at the nominal focal plane - characterised by either a change in beam width or a reduction in the focused intensity. A related study on accurate simulations of tilted circular zone plates (ZP) for x-rays was presented by Ali \& Jacobsen in \cite{Sajid2020}. However, we show that their conclusions cannot be directly transposed to the case of x-ray refractive lenses.

In this paper, we studied experimentally the effect of rotating 1D and 2D lenses and compared the results with simulations done using the wave optical model in projection approximation presented in \cite{celestre_recent_2020,Barc4RO_oasysgit}. We measure and simulate 2D focusing beryllium lenses with $R=\SI{50}{\micro\meter}$ and 1D Be lenses with $R=\SI{100}{\micro\meter}$ from a commercial supplier \cite{rxoptics}. We find that tilting a 1D lens around its focusing direction can be used to fine-tune its focal length, which we show experimentally can be implemented in an x-ray beamline. 

\section{Methods}\label{sec:methods}

\subsection{Experiments}\label{sec:experiments}

\begin{figure}[t]
    \centering
    \includegraphics[width=\textwidth]{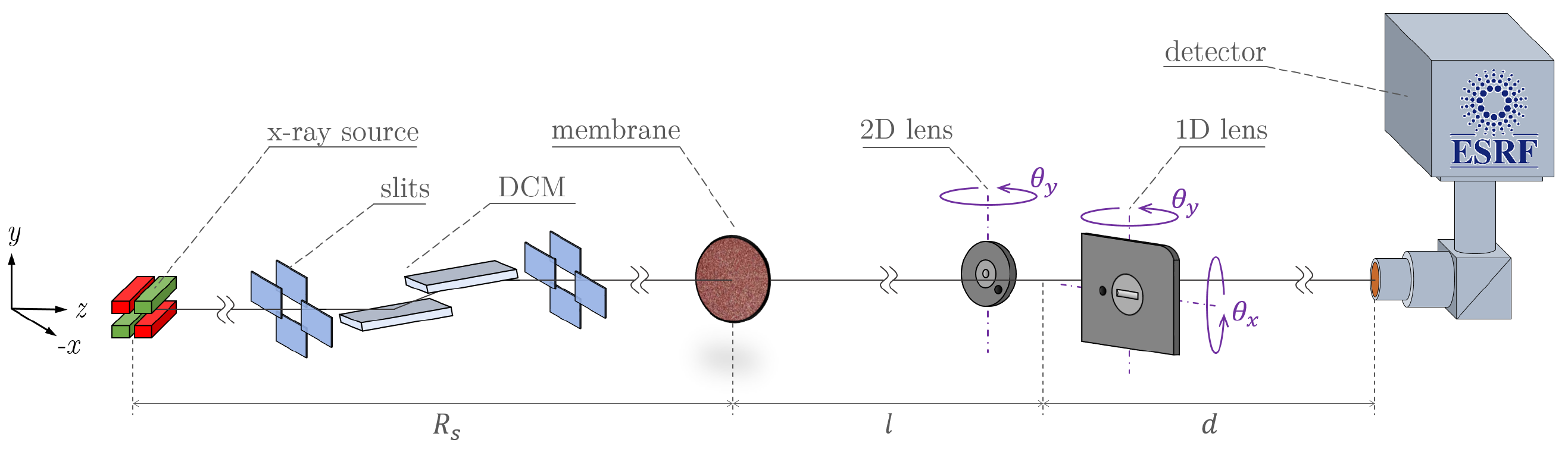}
    \caption{Schematic showing the experimental setup for XSVT metrology.}
    \label{fig:setup_sketch}
\end{figure}

\begin{figure}[t]
    \centering
    \includegraphics[width=\textwidth]{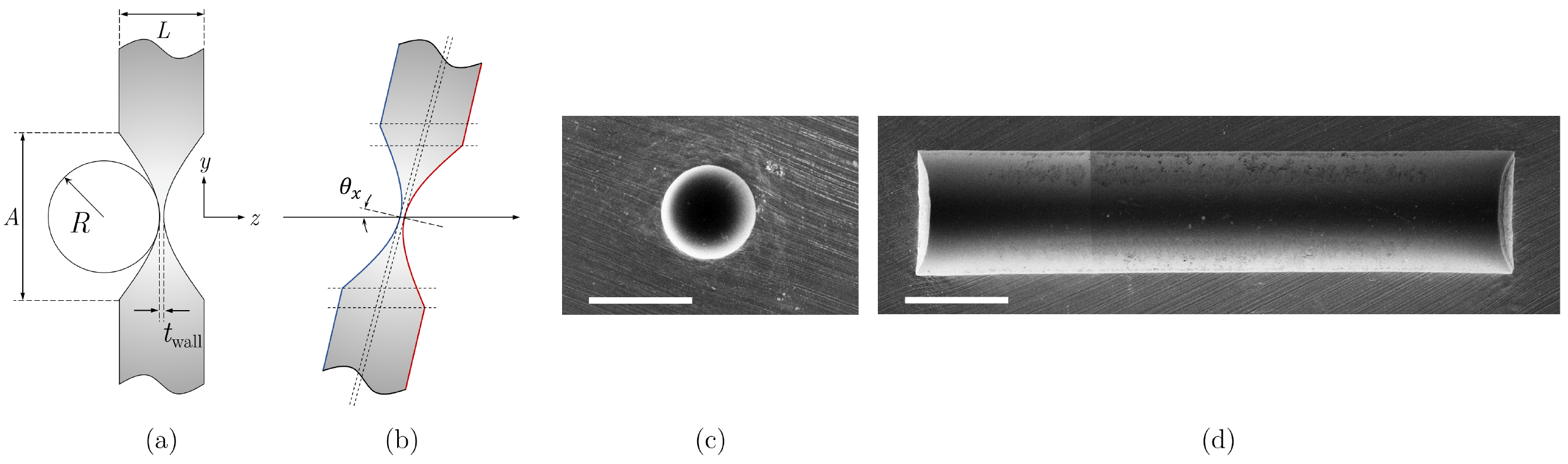}
    \caption{(a) X-ray lens main parameters. (b) tilted lens: the blue lines indicate the lens's front surface and the red lines indicate the back surface. Scanning electron microscope (SEM) image of a (c) 2D focusing Be lens with $R=\SI{50}{\micro\meter}$ and (d) of a 1D (vertical) focusing Be lens with $R=\SI{100}{\micro\meter}$. The white bars represent $\SI{500}{\micro\meter}$.}
    \label{fig:xray_lens}
\end{figure}

 The experimental data in the following section was obtained using x-ray speckle vector tracking (XSVT) in the differential metrology mode \cite{berujon_theory_2020,berujon_experiments_2020} --- a sketch of the setup is shown in Fig.~\ref{fig:setup_sketch}. This technique was chosen because it permits measurement of the phase gradients of the x-ray wavefront introduced by the lens, from which effects on the focusing behaviour can be inferred. The experiments were conducted at the ESRF-EBS BM05 beamline \cite{ziegler_esrf_2004} using a monochromatic beam at 17~keV (2D lens measurement) and 17.035~keV (1D lens measurement) employing a Si(111) double crystal monochromator (DCM) --- the difference in energies comes from the calibration of the DCM between the two experiments. The speckle modulator was placed  40.2~m away from the source; the lens was installed at 0.4~m downstream of the modulator and the detector, pco.edge Gold, 0.8~m downstream of the lens. Imaging was done using a $\SI{17}{\micro\meter}$ thick GGG:Eu scintillator coupled to a $10\times$ microscope objective (effective pixel size of $\sim\SI{0.635}{\micro\meter}$) for measurements of the 2D lens and to a $4\times$ magnification (effective pixel size of $\sim\SI{1.578}{\micro\meter}$) for the 1D lens due to its larger horizontal geometric aperture --- see Fig.~\ref{fig:xray_lens}.

\subsection{Modelling tilted x-ray lenses}\label{sec:modelling}

In this work, modelling uses physical optics descriptions \cite{Goodman_book}, that is, lenses are represented as a transmission element \cite{Cloetens_1996, Paganin_book} compatible with the scalar-diffraction theory used in many x-ray beamline design software packages. Simulations of x-ray lenses including misalignment and fabrication errors were initially presented in \cite{celestre_recent_2020} with an accompanying Python library in \cite{Barc4RO_oasysgit} to be used transparently with SRW \cite{codeSRW}, a standard software package for (partially-) coherent beamline simulation. This representation of x-ray lenses is generic enough to be used with any x-ray beamline wave-optics design package.

Take a Cartesian coordinate system $(x,y,z)$ in which the $z$ indicates the optical axis. A general phase object interacts with an electric field $E_\omega$ as:
\begin{equation}
     E_\omega(x,y,z^+)= T_\omega(x,y) \cdot E_\omega(x,y,z^-),
\end{equation}
where $E_\omega(x,y,z^-)$ is the electric field immediately before said transmission element, $E_\omega(x,y,z^+)$ is the transmitted wave-field and $T_\omega(x,y)$ is the complex transmission element in projection approximation:
\begin{equation}\label{eq:T}
     \mathrm{T}_\omega(x,y)=\exp\big\{-ik\Delta_z(x,y)\big[\delta_\omega(x,y)-i\beta_\omega(x,y) \big]\big\},
\end{equation}
with $k=2\pi\big/\lambda$ being the wavenumber, $n_\omega=1-\delta_\omega+i\cdot\beta_\omega$, the index of refraction in the x-ray regime and $\Delta_z(x,y)$ is the projected thickness along the optical axis direction --- see \S2.2 in \cite{Paganin_book}. The subscript $\omega$ is used as a reminder of the energy dependency of the generic transmission element. The real part of Eq.~\ref{eq:T}, that is, $\mathcal{R}\{\mathrm{T}_\omega(x,y)\}$ expresses the amplitude absorption, while the imaginary part $\mathcal{I}\{\mathrm{T}_\omega(x,y)\}$ describes the modification of the optical phase by the element $\varphi(x,y)$ and optical path length $\mathcal{W}(x,y)$ by:
\begin{equation}\label{eq:wave}
k\Delta_z(x,y)\delta_\omega=\varphi(x,y)=k\mathcal{W}(x,y).
\end{equation} 
A perfect (ideal) parabolic bi-concave x-ray lens centred around the transverse coordinates $(\Delta x,\Delta y)$ has the projected thickness $\Delta_z(x,y)$ given by:
\begin{equation}\label{eq:ProjecThick}
    \Delta_z(x,y) = 
        \begin{cases}
      \cfrac{(x-\Delta x)^2}{R_x}+\cfrac{(y-\Delta y)^2}{R_y}+t_\text{wall}, &\forall~(x-\Delta x,y-\Delta y) \in A,\\
      L, &\text{otherwise}.
        \end{cases}
\end{equation}  
$R_x$ and $R_y$ are the radii of curvature at the apices of the parabolic sections in the horizontal and vertical directions;  $t_{\text{wall}}$ is the distance between parabolic sections at the apices, $L$ is the lens thickness; and $A$ is the lens geometric aperture. An image of a lens cross-section is depicted in Fig.~\ref{fig:xray_lens}(a). The projected thickness described in Eq.~\ref{eq:ProjecThick} can be substituted into Eq.~\ref{eq:T} to represent the transmission element associated with an ideal x-ray lens. The focal length associated with each refracting surface is given by $f_{x,y}=R_{x,y}\big/\delta_\omega$, and if $R_x=\infty$ or $R_y = \infty$, there is no focus on this direction. In the thin lens approximation a lens with $2N$ refracting surfaces has $f_{x,y}=R_{x,y}\big/2N\delta_\omega$. The projected thickness of an ideal lens rotated around the $x-$ or/and the $y-$axis can be obtained by: i) individually generating a 3D point cloud for both front- and back surfaces of the lenslet: $z_\text{fs}(x,y)$ and $z_\text{bs}(x,y)$, where: $-z_\text{fs}(x,y)=z_\text{bs}(x,y)=\Delta_z(x,y)\big/2$; ii) applying the rotation matrices \cite{House2016} to each point in the cloud:
\begin{align}\label{eq:affine}
    R_{x}^\theta= \begin{bmatrix}
                    1 & 0 & 0 \\
                    0 & c_\theta & -s_\theta \\
                    0 & s_\theta & c_\theta 
    \end{bmatrix} ,\quad
    R_{y}^\theta = \begin{bmatrix}
                    c_\theta & 0 & s_\theta\\
                    0 & 1 & 0 \\
                    -s_\theta & 0 & c_\theta
    \end{bmatrix} ,\quad
    R_{z}^\theta = \begin{bmatrix}
                c_\theta & -s_\theta & 0\\
                s_\theta &c_\theta & 0 \\
                0 & 0 & 1
\end{bmatrix},
\end{align}
with $c_\theta=\cos(\theta)$ and $s_\theta=\sin(\theta)$; $R_{x}^\theta$ denotes a rotation around the $x-$axis, $R_{y}^\theta$ around the $y-$axis; iii) re-deposition (interpolation) of the rotated point clouds $z^{R^\theta}_\text{fs}(x,y)$ and $z^{R^\theta}_\text{bs}(x,y)$ on the same $(x,y)$ grid. Finally, iv) the projected thickness can be obtained by:
\begin{equation}\label{eq:ProjecThick_R}
    \Delta^{R^\theta}_z(x,y) = z^{R^\theta}_\text{bs}(x,y) - z^{R^\theta}_\text{fs}(x,y),
\end{equation}  
which can be directly substituted for $\Delta_z(x,y)$ in Eq.~\ref{eq:T}. A tilted lens is depicted in Fig.~\ref{fig:xray_lens}(b).

\subsubsection{Simulations}\label{sec:simulations}

X-ray intensity distributions are generated using SRW by propagating a wavefront 800~mm downstream from the transmission element representing the lens, which is implemented using the aforementioned CRL model - this is available through \texttt{barc4ro} \cite{Barc4RO_oasysgit}. The optical elements are illuminated by a monochromatic and coherent Gaussian wavefront at 17~keV for the 2D lenses and 17.035~keV for the 1D lenses. The phase gradients are calculated by generating the thickness profile (also using using \texttt{barc4ro}), multiplying it by $\delta_{\text{Be @ 17~keV}}=1.178476\cdot10^{-6}$ or  $\delta_{\text{Be @ 17.035~keV}}=1.173637\cdot10^{-6}$ (for the 1D case) in order to obtain the phase and then, applying the \texttt{Numpy} function \texttt{gradient}~\cite{npGradient}. 

\section{Results}\label{sec:example_sim}

In this section we present two examples: the tilting of a 2D focusing Be lens with $R=\SI{50}{\micro\meter}$ (§\ref{sec:2Dlens}) and of a 1D (vertical) focusing Be lens with $R=\SI{100}{\micro\meter}$ (§\ref{sec:1Dlens}) both from a commercial supplier \cite{rxoptics}. As indicated in Fig.~\ref{fig:setup_sketch}, both lenses are rotated around the $y-$axis for a range of $\theta_y$ angles - the 1D lens is also rotated around $\theta_x$. Experimental data is accompanied by equivalent simulated data.

\subsection{2D Be lens}\label{sec:2Dlens}

\begin{figure}[t!]
    \centering
    \includegraphics[width=105mm]{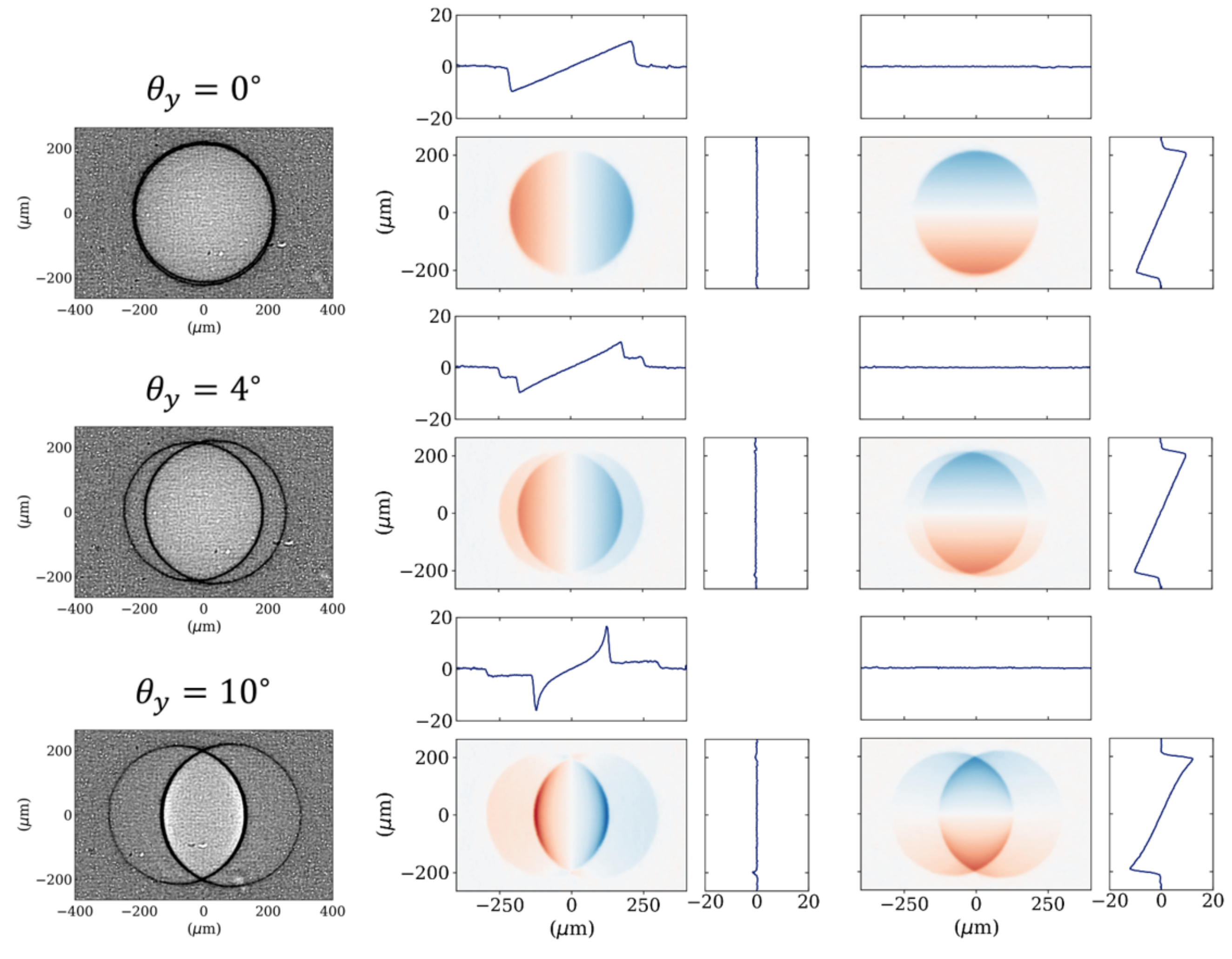}
    \caption{Experimental data for a 2D Be lens with $R=\SI{50}{\micro\meter}$ at $\text{E}=17$~keV. left column: intensity profiles (radiographs); central column: gradient along the $x-$direction (horizontal); right column: gradient along the $y-$direction (vertical). Gradients are in $\SI{}{\micro\rad}$. The profile cuts presented with the gradients are profiles passing through $(0, 0)$.}
    \label{fig:exp_Be_50um}
\end{figure}

\begin{figure}[t!]
    \centering
    \includegraphics[width=105mm]{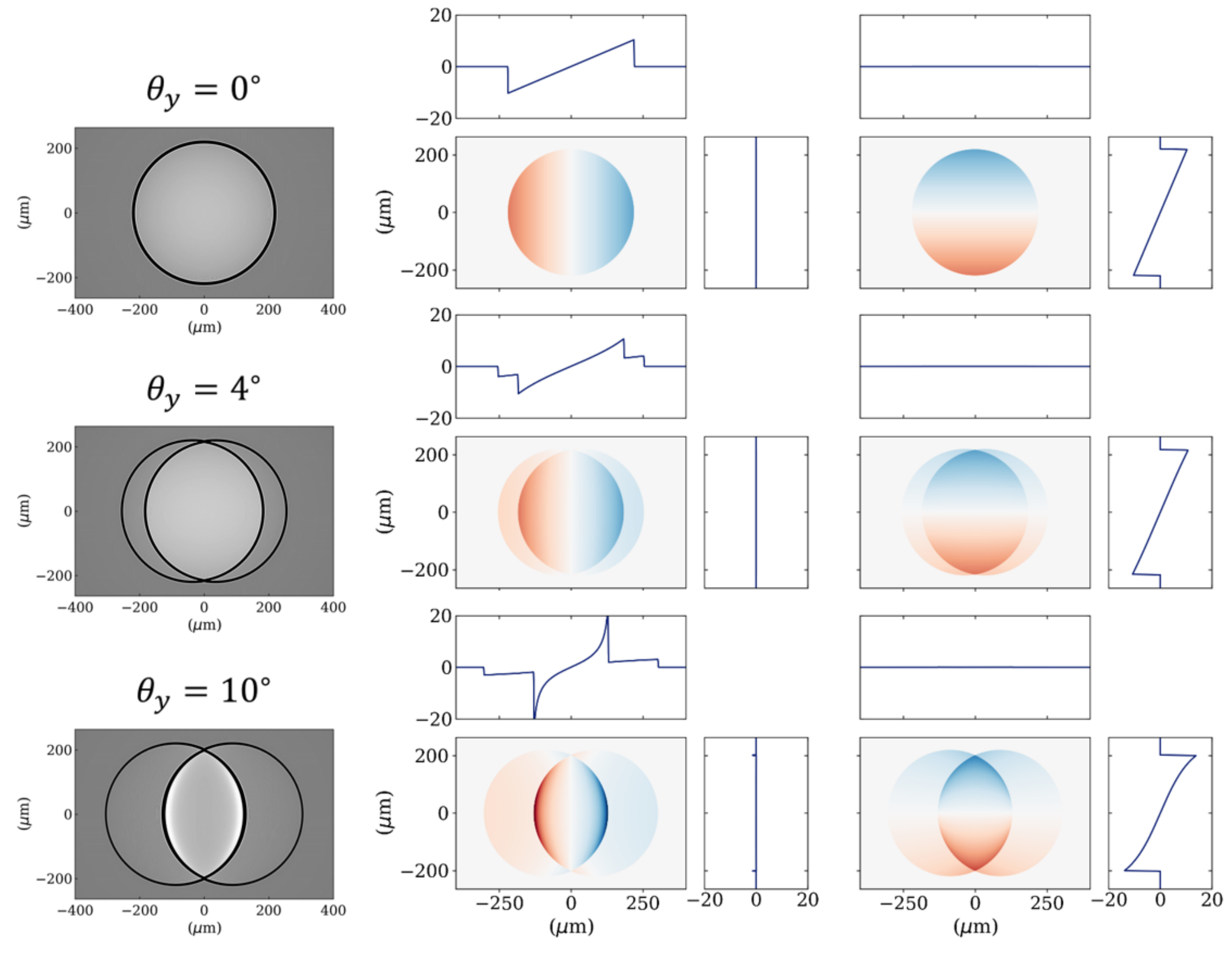}
    \caption{Simulation data for a 2D Be lens with $R=\SI{50}{\micro\meter}$ at $\text{E}=17$~keV. left column: intensity profiles (radiographs); central column: gradient along the $x-$direction (horizontal); right column: gradient along the $y-$direction (vertical). Gradients are in $\SI{}{\micro\rad}$. The profile cuts presented with the gradients are profiles passing through $(0, 0)$.}
    \label{fig:sim_Be_50um}
\end{figure}

The horizontal and vertical experimental phase gradients of a 2D bi-concave beryllium lens with $R=\SI{50}{\micro\meter}$ are presented in Fig.~\ref{fig:exp_Be_50um} and the equivalent simulated data in Fig.~\ref{fig:sim_Be_50um}. Such a lens typically is $L=1$~mm thick with $t_\text{wall}\approx\SI{35}{\micro\meter}$ and geometric aperture of $A\approx\SI{440}{\micro\meter}$. These parameters are the ones used for the simulations and it is assumed that the measured lens follows them closely.

For the perfectly aligned lens, i.e. no tilt, both horizontal and vertical gradients are linear --- this is expected as this lens has a quadratic thickness profile (see Eq.~\ref{eq:ProjecThick}). As soon as tilts start to be introduced, a reduction in the lens's effective geometric aperture due to the projected overlap of the paraboloidal-profiled front- and back-refracting surfaces [see Fig.~\ref{fig:xray_lens}(b)] becomes evident even at small angles. The gradients also lose their linearity, converging to odd-polynomial shapes of a higher order --- this change is smooth and less noticeable for smaller angles. Both these effects are present in the direction of the tilt and orthogonal to it, albeit weaker in the orthogonal direction.

\begin{figure}[t!]
    \centering
    \includegraphics[width=103mm]{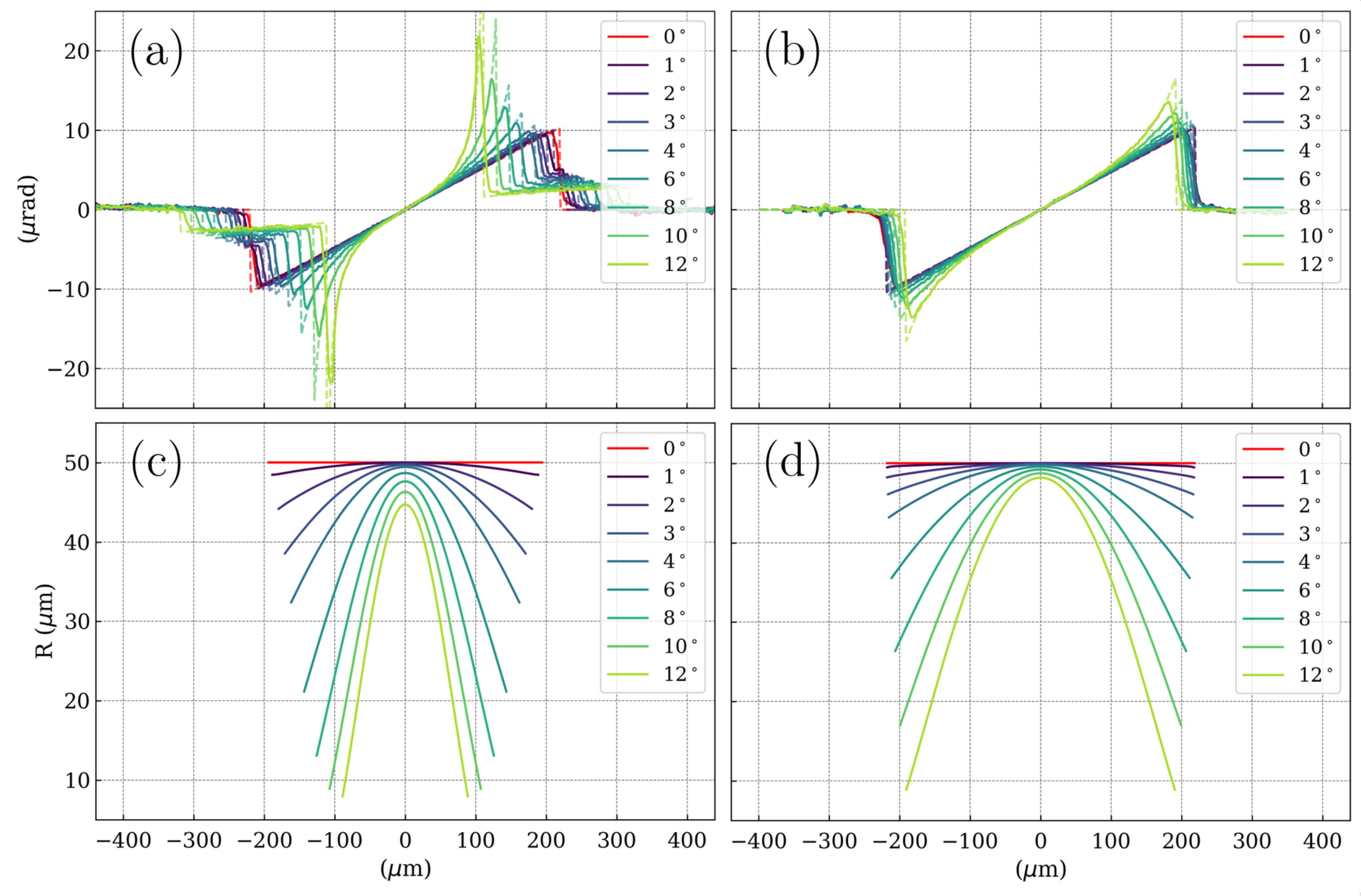}
    \caption{(a) horizontal and (b) vertical gradient cuts comparison for the simulation (dashed line) and experimental data (continuous line) for a 2D Be lens with $R=\SI{50}{\micro\meter}$ at $\text{E}=17$~keV. Calculated (c) horizontal and (d) vertical apparent radius of curvature.}
    \label{fig:com_sim_be50um}
\end{figure}

The qualitative agreement between experimental data (Fig.~\ref{fig:exp_Be_50um}) and simulations (Fig.~\ref{fig:sim_Be_50um}) is very good, despite some small asymmetries seen in the experimental radiographs (for lower angles). These could be related to fabrication misalignments or the position of the pivot point for the rotation. A direct superposition of the measured gradients is shown in Fig.~\ref{fig:com_sim_be50um}(a) and (b): continuous lines represent cuts in the experimental data and dashed lines represent simulation data. This figure confirms quantitatively the already seen qualitative agreement between modelled and measured lenses. However, the measured gradients do not present the same steep discontinuity at the edges of the effective geometric aperture. Two possible effects might explain this behaviour: the transition between the parabolic section and the flat area surrounding the lens active area is not sharply defined experimentally. These are embossed lenses and during their production, excess material is pushed away to the sides --- this is very clear in Fig.~\ref{fig:xray_lens}(c). A second reason may be related to the experimental technique and its sensitivity: the areas where the transition from parabolic to flat surfaces occurs generate strong phase contrast --- see black circles (radiographs) in Figs.~\ref{fig:exp_Be_50um} and Fig.~\ref{fig:sim_Be_50um}. The speckle signal in these areas has lower visibility and the data quality is consequently lower. 

\newpage
To infer the effects of tilting on an x-ray beam, we calculate the local radius of curvature from vertical and horizontal phase gradient profiles extracted from the simulated data:
\begin{equation}\label{eq:R_exp}
   R_{x,y}=\cfrac{2\cdot\delta_\omega}{\nabla^2\mathcal{W}}.
\end{equation}
The first derivative of the wavefront phase (or optical path difference --- see Eq.~\ref{eq:wave}) gives its slopes. The linear approximation of the slopes gives the average curvature of the two surfaces of the lens superimposed at ${x,y}$, that is, $1\big/R_{x,y}$.
The results are shown in Fig.~\ref{fig:com_sim_be50um}(c) and (d). Generally, what is observed for both horizontal and vertical directions is a smooth variation of the radius of curvature over the geometric aperture with a smaller $R$ towards the edges --- contrary to what one may be led to believe if the fit were to be done just in the central part of the lens. This variation in $R$ means that the outer parts of the lens will focus x-rays upstream of the x-rays focused by the central region, thus creating a $\sminus$\!$\prec$ shaped beam caustic similarly to what has been reported in \cite{schropp_2013,Seiboth2018,celestre_modelling_2020}. This is typical of spherical aberrations (first and higher orders). The difference between the $R$ variation in the horizontal and vertical directions also gives rise to some astigmatism. 

\subsection*{The case of the plane-concave 2D Be lens}\label{sec:2Dlens_single}

\begin{figure}[t!]
    \centering
    \includegraphics[width=\textwidth]{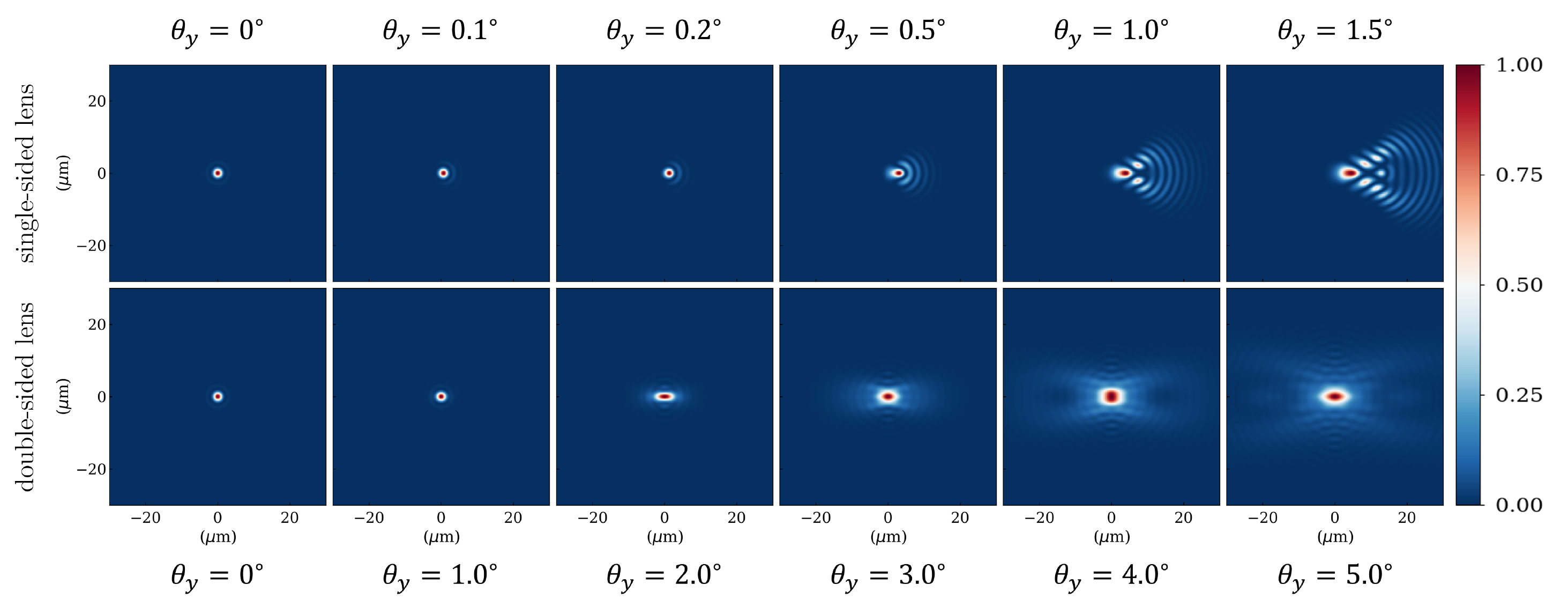}
    \caption{PSF calculations for single- and double-sided lenses with the same focal length at $\text{E}=8$~keV.}
    \label{fig:single-sided_lens}
\end{figure}

As indicated in the \nameref{sec:intro} (§\ref{sec:intro}), related work on tilted zone plates (ZP) has been reported by Ali \& Jacobsen in \cite{Sajid2020}. Among other results, they show that moderate tilts on a zone plate will produce a point spread function (PSF) typical of coma aberration (cf.~$Q=0.33$ in Fig.~5 ibid.) or astigmatism (cf.~$Q=0.33$ in Fig.~6 ibid.) depending on the energy the zone plates are employed. This is not the case for refractive x-ray lenses. Regardless of the energy at which it is used, a 2D plane-concave lens displays a PSF typical of coma when tilted. This is because it has only one parabolic refracting surface (behaving similarly to the lower energy ZP in terms of the PSF). A 2D bi-concave lens, however, presents a mix of spherical aberration and astigmatism as discussed previously --- behaviour that is also seen in the higher energy ZP in the image plane. Figure~\ref{fig:single-sided_lens} presents the simulated PSF at 8~keV for a bi-concave 2D Be lens with $R=\SI{50}{\micro\meter}$ ($\delta_{\text{Be @ 8.0~keV}}=5.326451\cdot10^{-6}$) and a hypothetical plane-concave lens also with $R=\SI{50}{\micro\meter}$, but with a $\delta_{\text{p-c}}=2\cdot\delta_{\text{Be @ 8.0~keV}}$ in order for both lenses to have the same numerical aperture. 

\subsection{1D Be lens}\label{sec:1Dlens}

When describing tilted 1D lenses, an important distinction has to be done regarding the direction of rotation. Take the lens depicted in Fig.~\ref{fig:xray_lens}(d): it has a vertical parabolic profile with an associated $R_y$ and a flat horizontal profile with $R_x=\infty$, thus this lens will focus vertically in the $y-$direction. For this particular lens and orientation, a tilt around the focusing direction will be represented by $\theta_y$ and will be referred to as horizontal tilt, while rocking along the focusing direction, that is $\theta_x$, will be called vertical tilt. In this section, we model and measure a commercial 1D bi-concave Be lens with $R_y\approx\SI{93}{\micro\meter}$, $L=1$~mm, $A_x\approx2.7$~mm and $A_y\approx\SI{620}{\micro\meter}$. This optical element should correspond to a nominal lens radius of $R=\SI{100}{\micro\meter}$, but our metrology yields a different value for $R$, which is the value we adopt for simulations to facilitate direct comparison.

\subsection{Vertical tilt (rotation $\theta_x$)}

\begin{figure}[t!]
    \centering
    \includegraphics[width=120mm]{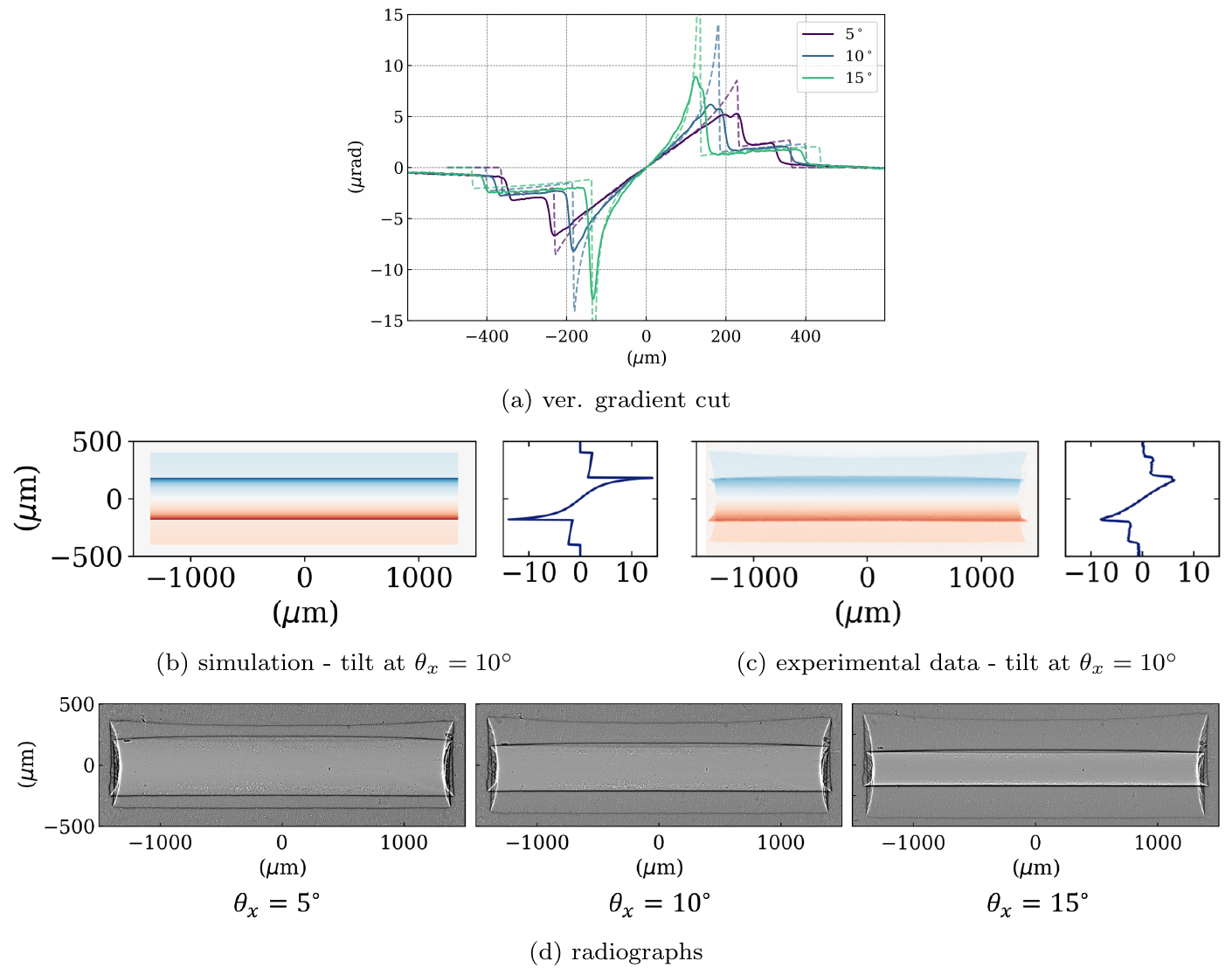}
    \caption{Tilt along the focusing direction. (a) vertical gradient cuts comparison for the simulation (dashed line) and experimental data (continuous line) for a 1D Be lens with $R\approx\SI{93}{\micro\meter}$ at $\text{E}=17.035$~keV. (b) simulated and (c) experimental vertical gradients for the said lens at $\theta_x=10^\circ$. (d) radiographs were taken during the tilt scan.}
    \label{fig:1D_Be_100um_orthogonal}
\end{figure}

For a vertically focusing lens, a tilt around $\theta_x$ will behave very similarly to a tilted 2D lens in the same direction and the observations in §\ref{sec:2Dlens} will apply. Perpendicular to the focusing direction, the gradient remains flat and no alterations are observed. Figure~\ref{fig:1D_Be_100um_orthogonal} shows simulated and experimental data for this case. Radiographs show that this lens in particular is a poor specimen and the edges of this lens are not sharp, straight nor overlap perfectly, which explains why the right side of the gradients in Fig.~\ref{fig:1D_Be_100um_orthogonal}(a) do not follow closely the simulations.

\subsection{Horizontal tilt (rotation $\theta_y$)}

\begin{figure}[t!]
    \centering
    \includegraphics[width=\textwidth]{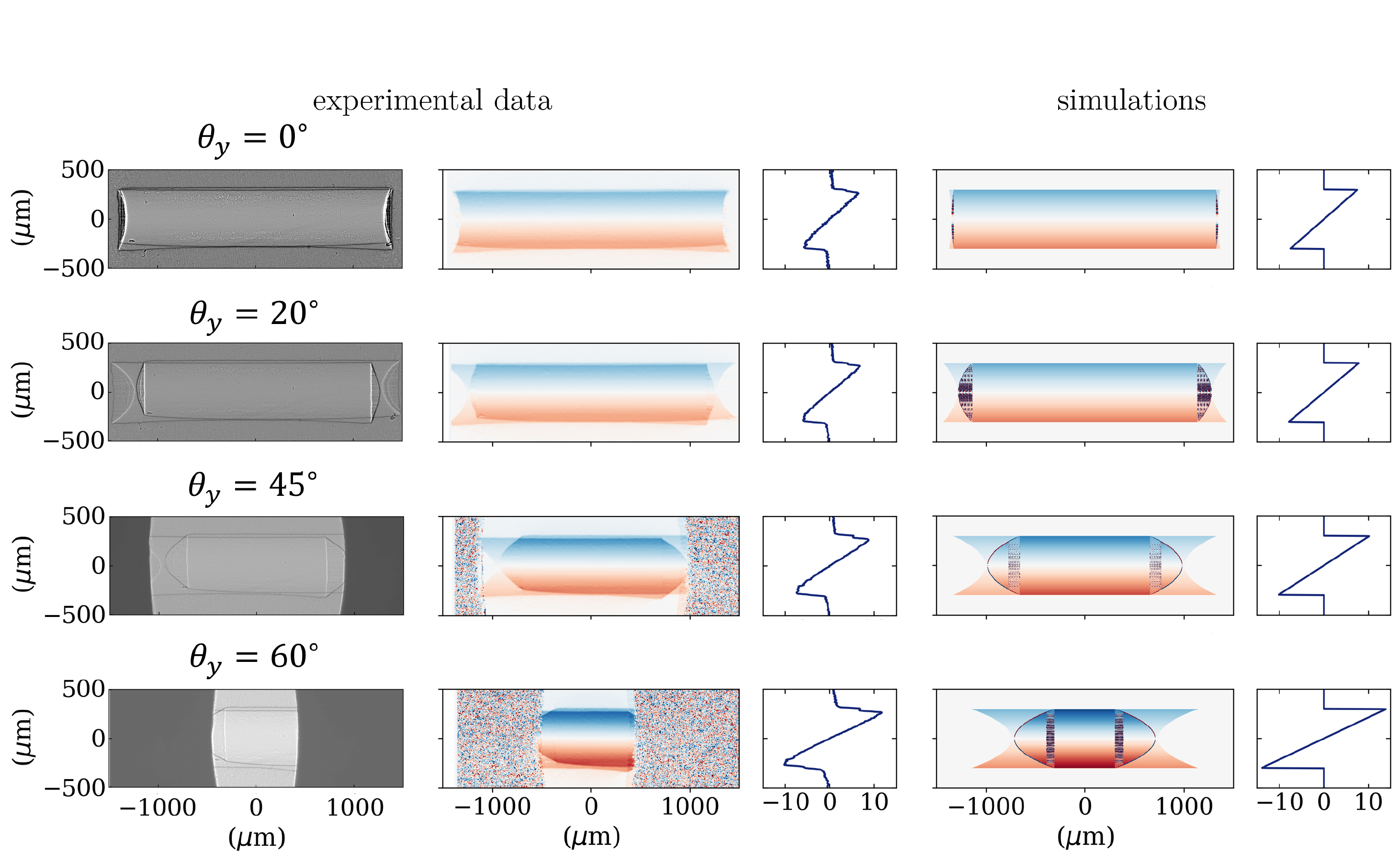}
    \caption{Experimental and simulated results for the angular scan of a 1D Be lens with $R\approx\SI{93}{\micro\meter}$ at $\text{E}=17.035$~keV.}
    \label{fig:1D_Be_tilt}
\end{figure}

In this section, we show that tilting a 1D lens around $\theta_y$ increases the projected thickness while still keeping a parabolic profile along the beam direction, which can be used to fine-tune its focusing. Figure~\ref{fig:1D_Be_tilt} shows radiographs and the vertical gradient for a 1D Be lens with $R\approx\SI{93}{\micro\meter}$ at $\text{E}=17.035$~keV. Once again the simulations can describe the mean features of experimental data: an increase in steepness of the vertical gradient and a reduction in the horizontal geometric aperture. The radiographs show shadowing caused by the metallic frame in which the Be lens is contained, which is not reproduced by the simulations --- this is the cause of the grainy areas in the experimental gradient (no speckle signal). The noise areas in the simulated gradients lie outside the reduced lens active area and are not a cause of concern. They come from the projection of material outside the active area into the lens geometric aperture. Unlike the tilt of a 2D lens, the horizontal gradients remain unaffected by the tilt.

\begin{figure}[t!]
    \centering
    \includegraphics[width=110mm]{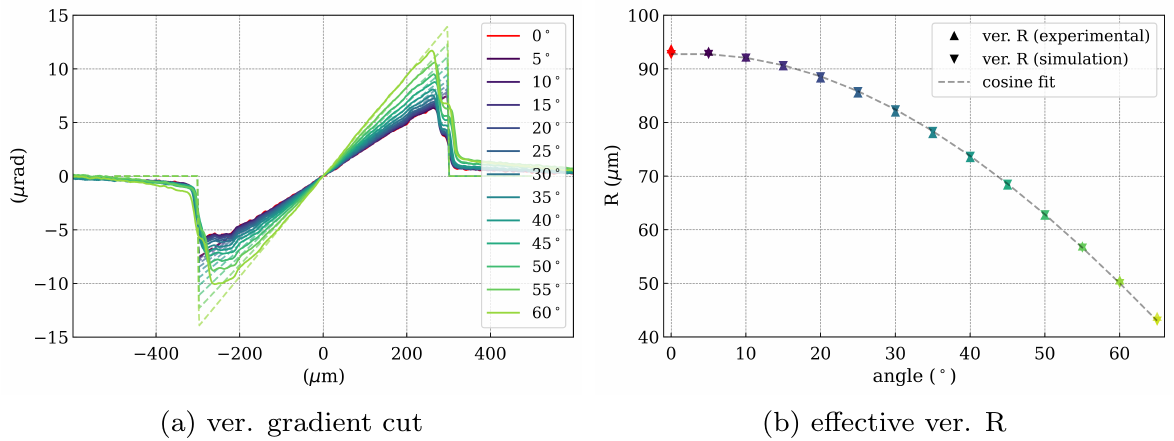}
    \caption{Experimental and simulated results for a angular scan of a 1D Be lens with $R\approx\SI{93}{\micro\meter}$ at $\text{E}=17.035$~keV.}
    \label{fig:comp_Be_100um}
\end{figure}

The plots of the vertical gradient cuts in Fig.~\ref{fig:comp_Be_100um}(a) show that not only do the gradients keep their linearity during the tilt scan but also their slopes follow closely those of the simulations. However, between $\SI{-300}{\micro\meter}$ and $\SI{-200}{\micro\meter}$ (lower part of the geometric aperture) experimentally-determined gradient data differs from the expected values (simulations) --- unlike what is observed between  $\SI{200}{\micro\meter}$ and $\SI{300}{\micro\meter}$, for example. The reasons for such mismatches are not connected to tilting or modelling but arise from the lens quality as discussed previously. 

The fact that the gradients maintain their linearity during the scan is a very important result as it indicates that such a tilted lens can be used in optical design without adding aberrations to the x-ray beam while focusing it. In this case, we can apply Eq.~\ref{eq:R_exp} and calculate the effective radius of curvature as a function of angle, which is shown in Fig.~\ref{fig:comp_Be_100um}(b). Both data sets show that the projected radius of curvature $R_\text{proj}$ follows very close a cosine function, which is the expected behaviour from an inclined cut of a parabolic cylinder: any oblique cut will give another parabola with a reduced apex radius. This allows us to model it as a function of $R$ and $\theta_y$:
\begin{equation}\label{eq:R_proj}
   R_\text{proj}=R_0\cdot\cos{(\theta_y-\theta_0)}=(92.8\pm0.2)\cdot\cos(\theta_y-2.61^\circ\pm0.04^\circ)~[\SI{}{\micro\meter}].
\end{equation}

\begin{figure}[t!]
    \centering
    \includegraphics[height=40mm]{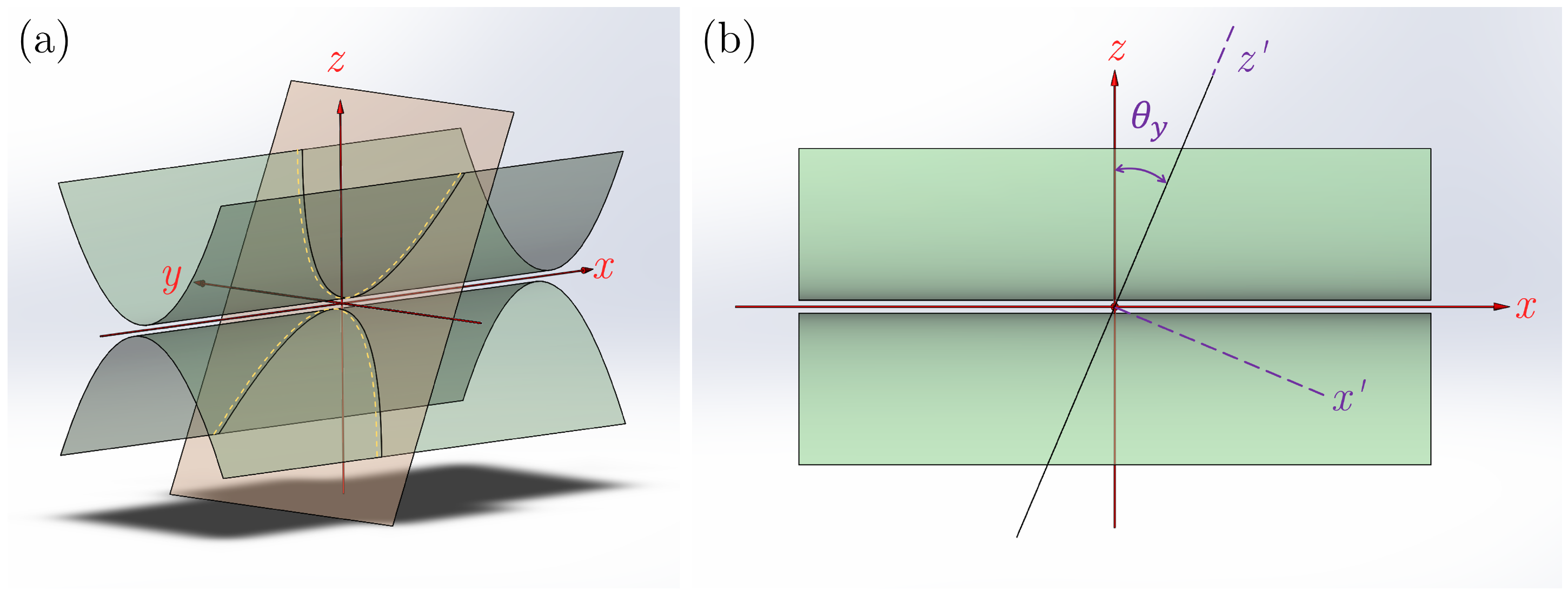}
    \caption{Oblique cut of a parabolic cylinder.}
    \label{fig:oblique_cut}
\end{figure}

The fit was obtained from non-linear least squares reduction of the experimental data in Fig.~\ref{fig:comp_Be_100um}(b) and it is plotted on the same figure (dashed line). We added a free parameter $\theta_0$ to the model accounting for any possible initial misalignment in the sampled lens. With the scan in Fig.~\ref{fig:comp_Be_100um}(b) we demonstrate experimentally that it is possible to use a lens with radius $R$ and continuously obtain any equivalent lens with a smaller radius of curvature down to and beyond $R_\text{proj}=R\big/2$, despite the accompanying reduction in the horizontal geometric aperture. Indeed, the behaviour in Eq.~\ref{eq:R_proj} is exactly what is expected analytically from an inclined cut of a parabolic cylinder. Consider the upper parabolic shell in Fig.~\ref{fig:oblique_cut}(a) with equation given by:
\begin{equation}\label{eq:parabo_cylinder}
    \frac{y^2}{2R_y}+\frac{t_\text{wall}}{2}-z=0.
\end{equation}  
Now consider an inclined plane including the $y$ axis and cutting the cylindrical shell at $\theta=\theta_y$ in Fig.~\ref{fig:oblique_cut}. 
Transforming Eq.~\ref{eq:parabo_cylinder} to the new coordinate system $x'$, $y'$, $z'$ (with $y'$ parallel to $y$) defined by this plane and setting $x'= 0$, it is straightforward to show that the intersection of the parabolic shell with the plane is described by:
\begin{equation}\label{eq:parabolic_projection}
\frac{y'^2}{2\cos(\theta_y)R_y} +\frac{t_\text{wall}}{2\cos(\theta_y)}-z'=0.
\end{equation}
Consequently a beam propagating parallel to the $z'$ axis encounters a parabolic surface with apex radius $R_y\cos(\theta_y)$ which indeed describes the experimental behaviour (Eq. \ref{eq:R_proj}). 
The same reasoning applies to the lower parabolic shell in Fig.~\ref{fig:oblique_cut} and it is easy to show that analogous results are obtained for that surface.
\section{Exploiting tilted lenses: fine-tuning a CRL focusing at fixed energy}\label{sec:exploting}

For fixed energies, the focal length of a CRL, i.e.~a stack of $N$ lenslets, can only be adjusted by the addition or subtraction of individual lenses or sub-groups of those $N$ elements. This is the principle underlying the transfocator \cite{Vaughan2011}. Despite the density with which modern transfocators can be packed and the plurality of lenses they may contain \cite{Zozulya2012, Kornemann2017, Narikovich2019}, their accessible focal lengths still present discrete values. For sub-micron focusing applications, this discretisation does not impose practical limitations, since a large number of lenses implies a small percentage difference between focusing with $N$ and $N+1$ lenses. However, for moderate to weak focusing systems, this jump from $N$ to $N+1$ represents a rather large percentage change and it can be desirable to have finer control over it. 

Using the findings from the previous section, we demonstrate the feasibility of fine-tuning the focal length of a CRL with $N$ elements and making it work as if $N+1$ lenses were used. First, we define the optical power of a single lens as $F=f^{-1}=m\delta_\omega\big/R$; $m$ is the number of refracting surfaces: 2 for bi-concave, 1 otherwise. We can add several lenses in close contact, that is, neglecting the space between them, by $F_{\text{lenses}}=\sum F_i$. Consider a CRL stack of $N-1$ lenses of radius $R_1$ such that $F_1=m_1\delta_1 (N-1) \big/R_1$ and a single tiltable lens for fine-tuning $F_2=m_2\delta_2\big/R_{\text{proj}}$. The CRL will have a focal length given by:
\begin{equation}\label{eq:f_tilted}
    f_{\text{CRL}}=\cfrac{R_1R_2\cos{(\theta)}}{m_2\delta_2R_1 + m_1 \delta_1  (N-1) R_2\cos{(\theta)}},
\end{equation}  
which simplifies to:
\begin{equation}\label{eq:f_tilted_eq}
f_{\text{CRL}}=\cfrac{1}{2\delta_\omega}\cdot\cfrac{R}{N-1 + \cfrac{1}{\cos(\theta)}},
\end{equation}  
when using bi-concave lenses of the same type for fine-tuning, that is, $R_1=R_2=R$, $\delta_1=\delta_2=\delta_\omega$ and $m_1=m_2=2$.
It is easy to show that Eq.~\ref{eq:f_tilted_eq} reduces to $f_{\text{CRL}}\approx R\big/2\delta_\omega N$ when $\theta\approx0^\circ$ and that $f_{\text{CRL}}\approx R\big/2\delta_\omega (N+1)$ when $\theta$ approaches $60^\circ$.

\begin{figure}[t!]
    \centering
    \includegraphics[width=120mm]{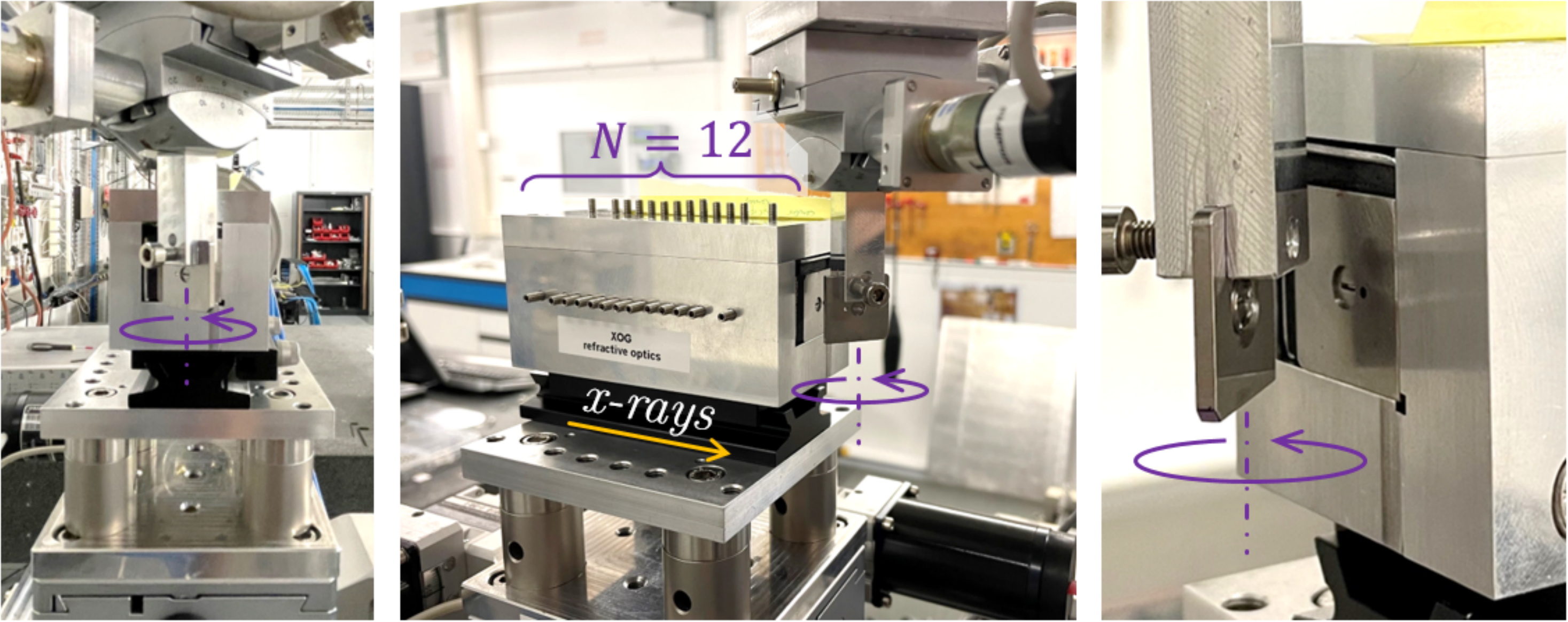}
    \caption{Experimental setup for fine-tuning the focal length of a CRL at fixed energy.}
    \label{fig:experimental_setup}
\end{figure}

\begin{figure}[t!]
    \centering
    \includegraphics[width=93mm]{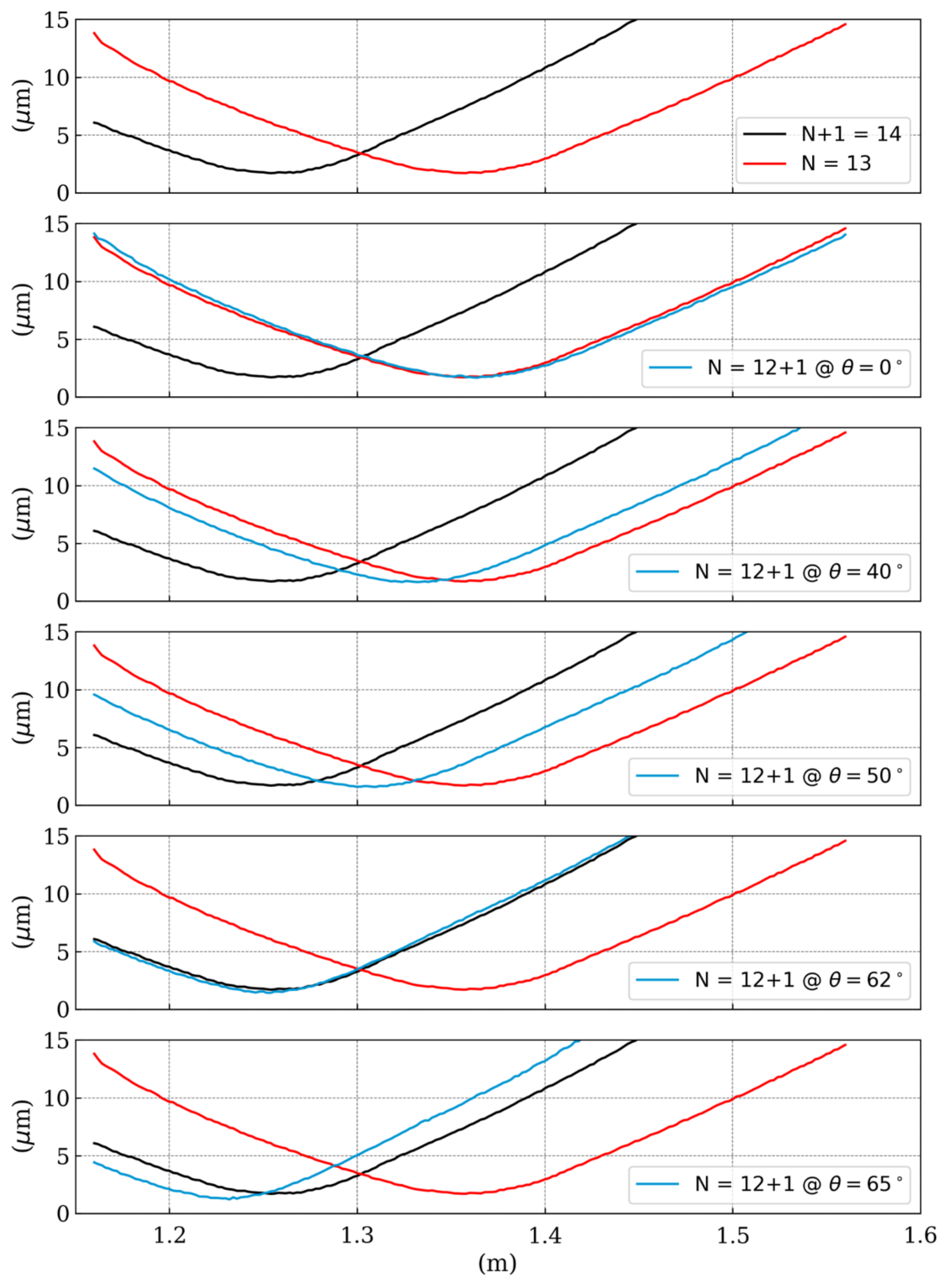}
    \caption{Experimental data: beam focusing as a function of tilt angle at $\text{E}=11.103$~keV. Red curves represent the upper limit imposed by $N$ lenses, black curves represent the lower limit imposed by $N+1$ lenses and the blue lines show the moving focal plane as a function of tilt.}
    \label{fig:1D_caustics}
\end{figure}

For this experiment, we used a stack of $N=13$ 1D Be lenses with $R=\SI{100}{\micro\meter}$ as the reference CRL. A second stack was created adding another lens of the same type to the first CRL to obtain $N+1=14$. Once the measurements using those stacks were done, we separated the $N=13$ pile into $N=12+1$ with the extra lens mounted outside the casing as shown in Fig.~\ref{fig:experimental_setup} --- this allowed us to align and rotate the lens around $\theta_y$ while keeping it centred with the stack. Upstream of the CRLs, a slit was placed cropping the beam down to about $\sim\SI{500}{\micro\meter}~\times~\SI{500}{\micro\meter}$, which was approximately the geometric aperture of a 1D Be lens with $R=\SI{100}{\micro\meter}$ tilted by $\theta_y\approx60^\circ$ (upper end of the tilt scan). The same detector used for the speckle tracking measurements (§\ref{sec:experiments}) was scanned along the optical axis to record the beam caustics, but this time with a magnification of $20\times$ (effective pixel size of $\sim\SI{0.32}{\micro\meter}$). The experiment was done using a monochromatic beam at $\text{E}=11.103$~keV.

\begin{figure}[t!]
    \centering
    \includegraphics[width=120mm]{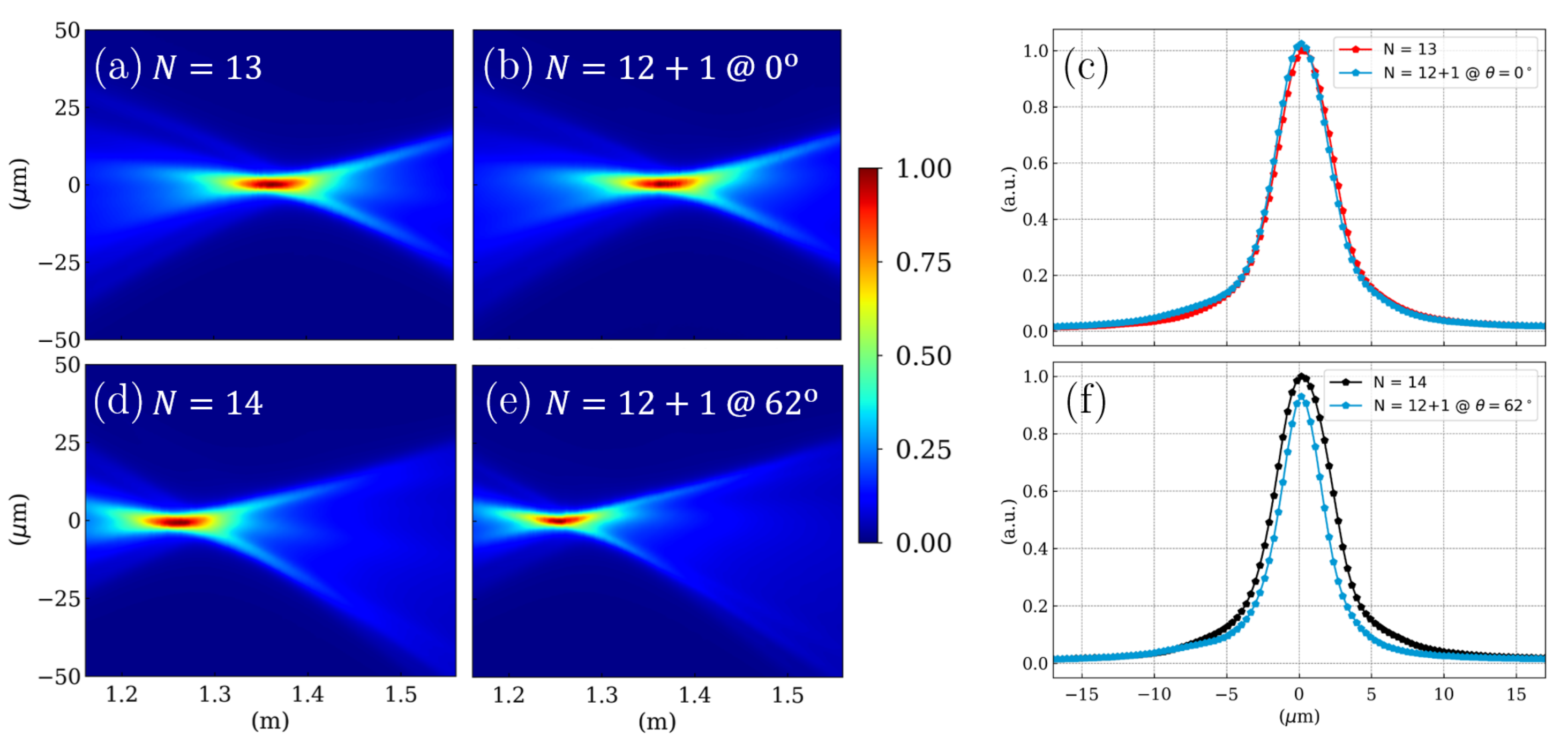}
    \caption{Experimental data: beam caustics at $\text{E}=11.103$~keV for (a) $N=13$ lenses, (b) $N=14$ lenses, (c) $N=12+1$ lenses with $\theta_y=0^\circ$ and (d) $N=12+1$ lenses with $\theta_y=60^\circ$. (e) and (f) beam profile at the focal plane.}
    \label{fig:caustics_2D_and_beamprofile}
\end{figure}

Figure~\ref{fig:1D_caustics} shows the full width at half maximum (FWHM) beam size as a function of the distance from the centre of the stacks. The top image shows the two limiting cases of $N$ (red) and $N+1$ (black) focusing along the optical axis. The remaining graphs in Fig.~\ref{fig:1D_caustics} show the focusing of the system in Fig.~\ref{fig:experimental_setup} for a few selected tilt angles (in blue). We show that when there is no tilt, the system composed of $N=12+1$ behaves similarly to the original $N=13$. As we start to tilt the lens. we see the focal plane move upstream until it matches the focal plane of the $N=14$ system, which happens at $\theta_y\approx62^\circ$. It is possible to tilt the lens even more and have an even shorter focal length as shown in the bottom of Fig.~\ref{fig:1D_caustics}, but the projected geometric aperture of the lens rapidly restricts the horizontal beam acceptance. The beam caustics for the two limiting cases and the tilted lenses emulating them are presented in Fig.~\ref{fig:caustics_2D_and_beamprofile}, along with the beam profiles at their respective focal planes. When comparing the pairs $N=13$ and $N=12+1~(0^\circ)$, we see that the beam caustics are very similar and both the beam intensity and width at the focal plane are very close --- see Fig.~\ref{fig:caustics_2D_and_beamprofile}(a)-(c). This was already expected. However, the pairs $N=14$ and $N=12+1~(60^\circ)$ are a little bit less similar probably due to some small issues with lens mounting, centring and alignment --- refer to Fig.~\ref{fig:caustics_2D_and_beamprofile}(d)-(f). We also notice that the beam is smaller for the $N=12+1~(60^\circ)$ case: $\SI{4.3}{\micro\meter}$ as opposed to $\SI{5.4}{\micro\meter}$ for $N=14$, which amounts to almost $20\%$ difference in beam width at the focal plane. The beam from the  $N=12+1~(60^\circ)$ configuration is also slightly less intense: about $95\%$ of the peak intensity --- this is expected due to two main factors: the horizontal aperture is smaller than the beam-defining slits; and more absorption in the tilted lens as the projected thickness is obtained by dividing Eq.~\ref{eq:ProjecThick} by $\cos(\theta)$; the first factor being the more preponderant. Despite the differences, all the caustic elements are present in both cases: the asymmetries up and downstream the focal plane; and the three-lobed structure on the right side (after the focus).

\section{Conclusion}\label{sec:conclusion}

In this paper, we studied tilted x-ray refractive lenses and demonstrated their behaviour experimentally. We revisited the modelling presented in \cite{celestre_recent_2020, Barc4RO_oasysgit} and benchmarked it against experimental data (speckle tracking) with excellent results: the models are accurate. This is indicated by Fig.~\ref{fig:com_sim_be50um}(a)-(b) and Fig.~\ref{fig:comp_Be_100um}(a), for example. They can be used in optical simulations for predicting the effects of tilted refractive optics illuminated by x-rays. We recall that the code in \cite{Barc4RO_oasysgit} is open access and can be used transparently with SRW. 

Based on our findings, tilting 2D focusing lenses does not seem to be interesting: both along and orthogonal to the focusing direction aberrations are added to the wavefront. For a bi-concave lens, spherical aberrations and astigmatism are introduced by tilting; for a plano-concave lens, tilting introduces coma as the main aberration. We also discussed the tilting of 1D lenses, which was grouped into two segments: tilts along and orthogonal to the focusing direction. The first group behaves very similarly to the 2D lenses with the same observations and conclusions being applicable. However, the tilt around the focusing direction is useful: we experimentally demonstrate that, despite the reduction in aperture, a lens tilted around the focusing direction can smoothly change its apparent $R$ down to and beyond $R\big/2$. This behaviour is in agreement with theory and follows a cosine function. The downside to rotating a lens around the focusing direction is a reduction in the lens geometric aperture, which for sufficiently large lenses may not be an issue. An additional point worth making is that by tilting the lens, we are varying the focusing parameters $(p, q)$ across the lens surface --- therefore there is a limit of applicability of this approach related to the depth of focus of the beam which should be sufficiently large so as to not degrade the focus due to this effect. This is the case when tilting a single lens element or of weakly focusing stacks/small numerical apertures.

 Fine-tuning the focal length of a CRL by adding a tilted lens was tested and we were able to demonstrate experimentally that a CRL composed of $N$ lenses can behave as if $N+1$ lenses were being used. This was shown by the measurement of the beam caustics yielding the beam size and intensity evolution along the optical axis. A comparison of the beam shape and intensity at the focal plane was also performed. There are some small differences between the caustic shape for $N+1$ lenses and the $N$ stack with a tilted lens. While a less intense beam is certainly due to a smaller aperture and increased projected thickness, differences in beam shape are probably attributed to small issues with lens mounting, centring and alignment in the transverse direction.

Besides the obvious applications a varying focal length has, we explicitly mention other possible uses of tilted lenses in experimental setups. Two crossed-tilted 1D lenses could be used to fine-tune the focal length and/or correct astigmatism of a 2D lens stack. For this application, the reduction in the aperture accompanying the tilting of lenses is not really a concern. Since it is possible to adjust the horizontal and vertical focal length independently, correcting asymmetric divergence of the beam e.g. due to a thermal bump on a white beam mirror or monochromator \cite{Brumund2021a,Brumund2021b} is also a possible application. Again, two crossed-tilted 1D lenses could be used to guarantee a focused beam onto a sample during small-range energy scans typical of extended x-ray absorption fine structure experiments (EXAFS). Finally, for 1D lenses with very large $R$, e.g. for collimation upstream of a monochromator, commercial lenses have only a small set of available radii. There, fine-tuning via a rotation seems the only way for obtaining the desired $R$.


\begin{backmatter}
\bmsection{Funding}
This project has received funding from the European Union’s Horizon 2020 Research and Innovation Programme under grant agreement N$^{\circ}$ 101007417 (NFFA-Europe Pilot Joint Activities -- NEP).

\bmsection{Acknowledgments}
The authors thank Christian David (PSI) for early exchanges. We acknowledge the ESRF for providing beamtime on BM05; Philip Cook and Luca Capasso for helping with the beamline setup. Special thanks to Luca Rebuffi (ANL-APS) for refactoring \texttt{barc4RefractiveOptics} into \texttt{barc4ro} and making it available within OASYS \cite{Barc4RO_oasysgit}. 

\bmsection{Disclosures}
The authors declare no conflicts of interest.

\bmsection{Data Availability}
Data underlying the results presented in this paper are publicly available at \cite{paperGit}. Supplementary data may be obtained from the authors upon reasonable request.

\end{backmatter}


\bibliography{sample}

\begin{thebibliography}{10}
\newcommand{\enquote}[1]{``#1''}

\bibitem{Suehiro1991}
S.~Suehiro, H.~Miyaji, and H.~Hayashi, \enquote{{Refractive lens for X-ray
  focus},} {\protect\JournalTitle{Nature}} \textbf{352}, 385 (1991).

\bibitem{Michette1991}
A.~G. Michette, \enquote{{No X-ray lens},} {\protect\JournalTitle{Nature}}
  \textbf{353}, 510 (1991).

\bibitem{Tomie1994}
T.~Tomie, \emph{X-ray lenses} (Japan Patent 6-045288, 1994).

\bibitem{Snigirev1996}
A.~Snigirev, V.~Kohn, I.~Snigireva, and B.~Lengeler, \enquote{{A compound
  refractive lens for focusing high-energy X-rays},}
  {\protect\JournalTitle{Nature}} \textbf{384}, 49 (1996).

\bibitem{Lengeler2001}
B.~Lengeler, C.~G. Schroer, B.~Benner, T.~F. Günzler, M.~Kuhlmann,
  J.~Tümmler, A.~S. Simionovici, M.~Drakopoulos, A.~Snigirev, and
  I.~Snigireva, \enquote{Parabolic refractive x-ray lenses: a breakthrough in
  x-ray optics,} {\protect\JournalTitle{Nucl. Instrum. Methods. Phys. Res. A}}
  \textbf{467-468}, 944--950 (2001).

\bibitem{Howells1993}
M.~R. Howells, \enquote{{Mirrors for Synchrotron-Radiation Beamlines},} in
  \emph{New Directions in Research with Third-Generation Soft X-Ray Synchrotron
  Radiation Sources,}  (NATO Advanced Study Institute, 1993), 1st ed.

\bibitem{David2001}
C.~David, B.~Nöhammer, and E.~Ziegler, \enquote{Wavelength tunable diffractive
  transmission lens for hard x rays,} {\protect\JournalTitle{Applied Physics
  Letters}} \textbf{79}, 1088--1090 (2001).

\bibitem{Yan2010}
H.~Yan, H.~C. Kang, R.~Conley, C.~Liu, A.~T. Macrander, G.~B. Stephenson, and
  J.~Maser, \enquote{Multilayer laue lens: A path toward one nanometer x-ray
  focusing,} {\protect\JournalTitle{X-Ray Optics and Instrumentation}}
  \textbf{2010}, 401854 (2010).

\bibitem{Cederstrom2000}
B.~{Cederstr{\"o}m}, R.~N. {Cahn}, M.~{Danielsson}, M.~{Lundqvist}, and D.~R.
  {Nygren}, \enquote{{Focusing hard X-rays with old LPs},}
  {\protect\JournalTitle{Nature}} \textbf{404}, 951 (2000).

\bibitem{Jark2019}
W.~Jark, A.~Opolka, A.~Cecilia, and A.~Last, \enquote{Zooming x-rays with a
  single rotation in x-ray prism zoom lenses (xpzl),}
  {\protect\JournalTitle{Opt. Express}} \textbf{27}, 16781--16790 (2019).

\bibitem{Alianelli2007}
L.~Alianelli, M.~{S{\'{a}}nchez del R{\'{i}}o}, and K.~Sawhney,
  \enquote{{Ray-tracing simulation of parabolic compound refractive lenses},}
  {\protect\JournalTitle{Spectrochimica Acta Part B}} \textbf{62}, 593--597
  (2007).

\bibitem{Baltser2011}
J.~Baltser, E.~Knudsen, A.~Vickery, O.~Chubar, A.~Snigirev, G.~Vaughan,
  R.~Feidenhans'l, and K.~Lefmann, \enquote{{Advanced simulations of x-ray beam
  propagation through CRL transfocators using ray-tracing and wavefront
  propagation methods},} {\protect\JournalTitle{Proceedings of SPIE}}
  \textbf{8141}, 814111 (2011).

\bibitem{SanchezdelRio2012}
M.~{Sanchez del Rio} and L.~Alianelli, \enquote{{Aspherical lens shapes for
  focusing synchrotron beams},} {\protect\JournalTitle{Journal of Synchrotron
  Radiation}} \textbf{19}, 366--374 (2012).

\bibitem{Osterhoff2017}
M.~Osterhoff, C.~Detlefs, and C.~Ferrero, \enquote{{Aberrations in compound
  refractive lens systems: analytical and numerical calculations},}
  {\protect\JournalTitle{Proceedings of SPIE}} \textbf{10388}, 21 (2017).

\bibitem{celestre_modelling_2020}
R.~Celestre, S.~Berujon, T.~Roth, M.~Sanchez~del Rio, and R.~Barrett,
  \enquote{Modelling phase imperfections in compound refractive lenses,}
  {\protect\JournalTitle{Journal of Synchrotron Radiation}} \textbf{27},
  305--318 (2020).

\bibitem{Andrejczuk2010}
A.~Andrejczuk, J.~Krzywi{\'{n}}ski, Y.~Sakurai, and M.~Itou, \enquote{{The role
  of single element errors in planar parabolic compound refractive lenses},}
  {\protect\JournalTitle{Journal of Synchrotron Radiation}} \textbf{17},
  616--623 (2010).

\bibitem{Sajid2020}
S.~Ali and C.~Jacobsen, \enquote{Effect of tilt on circular zone plate
  performance,} {\protect\JournalTitle{J. Opt. Soc. Am. A}} \textbf{37},
  374--383 (2020).

\bibitem{celestre_recent_2020}
R.~Celestre, O.~Chubar, T.~Roth, M.~Sanchez~del Rio, and R.~Barrett,
  \enquote{Recent developments in {X}-ray lens modelling with {SRW},}
  {\protect\JournalTitle{Proceedings of SPIE}} \textbf{11493}, 88--101 (2020).

\bibitem{Barc4RO_oasysgit}
R.~Celestre and L.~Rebuffi. Oasys-barc4ro {G}it{H}ub repository accessed in
  01/09/2022:\\\url{https://github.com/oasys-kit/oasys-barc4ro}.

\bibitem{rxoptics}
RXOPTICS GmbH \& Co. KG, Monschau, Germany.

\bibitem{berujon_theory_2020}
S.~Berujon, R.~Cojocaru, P.~Piault, R.~Celestre, T.~Roth, R.~Barrett, and
  E.~Ziegler, \enquote{X-ray optics and beam characterization using random
  modulation: theory,} {\protect\JournalTitle{Journal of Synchrotron
  Radiation}} \textbf{27}, 284--292 (2020).

\bibitem{berujon_experiments_2020}
S.~Berujon, R.~Cojocaru, P.~Piault, R.~Celestre, T.~Roth, R.~Barrett, and
  E.~Ziegler, \enquote{X-ray optics and beam characterization using random
  modulation: experiments,} {\protect\JournalTitle{Journal of Synchrotron
  Radiation}} \textbf{27}, 293--304 (2020).

\bibitem{ziegler_esrf_2004}
E.~Ziegler, J.~Hoszowska, T.~Bigault, L.~Peverini, J.~Y. Massonnat, and
  R.~Hustache, \enquote{The {ESRF BM05} metrology beamline: Instrumentation and
  performance upgrade,} {\protect\JournalTitle{AIP Conference Proceedings}}
  \textbf{705}, 436--439 (2004).

\bibitem{Goodman_book}
J.~W. Goodman, \emph{{Introduction to Fourier Optics}} (W. H. Freeman and
  Company, 2017), 4th ed.

\bibitem{Cloetens_1996}
P.~Cloetens, R.~Barrett, J.~Baruchel, J.-P. Guigay, and M.~Schlenker,
  \enquote{Phase objects in synchrotron radiation hard x-ray imaging,}
  {\protect\JournalTitle{Journal of Physics D}} \textbf{29}, 133--146 (1996).

\bibitem{Paganin_book}
D.~M. Paganin, \emph{Coherent X-Ray Optics} (Oxford University Press, 2006).

\bibitem{codeSRW}
O.~Chubar and P.~Elleaume, \enquote{Accurate and efficient computation of
  synchrotron radiation in the near field region,} {\protect\JournalTitle{Proc.
  EPAC-98}} pp. 1177--1179 (1998).

\bibitem{House2016}
D.~H. House and J.~C. Keyser, \emph{{Appendix C: Affine Transformations
  \textit{in} Foundations of Physically Based Modeling {\&} Animation}} (A K
  Peters/CRC Press, 2016), 1st ed.

\bibitem{npGradient}
\texttt{numpy.gradient} accessed in 05/08/2022:\\
  \url{https://numpy.org/doc/stable/reference/generated/numpy.gradient}.

\bibitem{schropp_2013}
A.~Schropp, R.~Hoppe, V.~Meier, J.~Patommel, F.~Seiboth, H.~J. Lee, B.~Nagler,
  E.~C. Galtier, B.~Arnold, U.~Zastrau, J.~B. Hastings, D.~Nilsson, F.~Uhlén,
  U.~Vogt, H.~M. Hertz, and C.~G. Schroer, \enquote{Full spatial
  characterization of a nanofocused x-ray free-electron laser beam by
  ptychographic imaging,} {\protect\JournalTitle{Scientific Reports}}
  \textbf{3} (2013).

\bibitem{Seiboth2018}
F.~Seiboth, F.~Wittwer, M.~Scholz, M.~Kahnt, M.~Seyrich, A.~Schropp, U.~Wagner,
  C.~Rau, J.~Garrevoet, G.~Falkenberg, and C.~G. Schroer,
  \enquote{{Nanofocusing with aberration-corrected rotationally~parabolic
  refractive X-ray lenses},} {\protect\JournalTitle{Journal of Synchrotron
  Radiation}} \textbf{25}, 108--115 (2018).

\bibitem{Vaughan2011}
G.~B.~M. Vaughan, J.~P. Wright, A.~Bytchkov, M.~Rossat, H.~Gleyzolle,
  I.~Snigireva, and A.~Snigirev, \enquote{{X-ray transfocators: focusing
  devices based on compound refractive lenses},} {\protect\JournalTitle{Journal
  of Synchrotron Radiation}} \textbf{18}, 125--133 (2011).

\bibitem{Zozulya2012}
A.~V. Zozulya, S.~Bondarenko, A.~Schavkan, F.~Westermeier, G.~Gr\"{u}bel, and
  M.~Sprung, \enquote{Microfocusing transfocator for 1{D} and 2{D} compound
  refractive lenses,} {\protect\JournalTitle{Optics Express}} \textbf{20},
  18967--18976 (2012).

\bibitem{Kornemann2017}
E.~Kornemann, O.~M\'{a}rkus, A.~Opolka, T.~Zhou, I.~Greving, M.~Storm,
  C.~Krywka, A.~Last, and J.~Mohr, \enquote{Miniaturized compound refractive
  x-ray zoom lens,} {\protect\JournalTitle{Optics Express}} \textbf{25},
  22455--22466 (2017).

\bibitem{Narikovich2019}
A.~Narikovich, M.~Polikarpov, A.~Barannikov, N.~Klimova, A.~Lushnikov,
  I.~Lyatun, G.~Bourenkov, D.~Zverev, I.~Panormov, A.~Sinitsyn, I.~Snigireva,
  and A.~Snigirev, \enquote{{CRL-based ultra-compact transfocator for X-ray
  focusing and microscopy},} {\protect\JournalTitle{Journal of Synchrotron
  Radiation}} \textbf{26}, 1208--1212 (2019).

\bibitem{Brumund2021a}
P.~Brumund, J.~Reyes-Herrera, C.~Detlefs, C.~Morawe, M.~Sanchez~del Rio, and
  A.~I. Chumakov, \enquote{{Design simulations of a horizontally deflecting
  high-heat-load monochromator},} {\protect\JournalTitle{Journal of Synchrotron
  Radiation}} \textbf{28}, 91--103 (2021).

\bibitem{Brumund2021b}
P.~Brumund, J.~Reyes-Herrera, C.~Morawe, T.~Dufrane, H.~Isern, T.~Brochard,
  M.~S{\'{a}}nchez~del R{\'\i}o, and C.~Detlefs, \enquote{{Thermal optimization
  of a high-heat-load double-multilayer monochromator},}
  {\protect\JournalTitle{Journal of Synchrotron Radiation}} \textbf{28},
  1423--1436 (2021).

\bibitem{paperGit}
Data underlying the results in this paper are available at:\\
  \url{https://gitlab.esrf.fr/celestre/tilted_lens}.

\end{thebibliography}

\end{document}